\documentstyle{article}

\begin{document}

\title{Sea-Boson Analysis of \\ the Infinite-U Hubbard Model}
\author{Girish S. Setlur \\ Harish Chandra Research Institute
 \\ Jhunsi, Allahabad }
\maketitle

\begin{abstract}
 By expanding the projection operator in powers of the density
 fluctuations, we conjecture a hamiltonian purely quadratic in the sea-bosons
 that reproduces the right spin and charge velocities and exponent for the 
 $ U = \infty $ case in one dimension known from the work of Schulz.
 Then we argue that by simply promoting wavenumbers
 to wave vectors we are able to study
 the two dimensional case. We find that the quasiparticle residue takes a
 value $ Z_{F} = 0.79 $ close to half-filling where it is the smallest.
 This is in exact agreement with the prediction by Castro-Neto and Fradkin
 nearly ten years ago. We also compute the magnetic suceptibility and find
 that it diverges close to half-filling consistent with Nagakoka's theorem.
\end{abstract}

\section{ Introduction }

 The large-U Hubbard Model or the t-J model and its variants have been the
 subject of active study ever since its importance to understanding cuprates
 has been realised.
 For a review see the article by Dagotto\cite{Dag}.
The slave-boson/fermion approach that naturally takes into account 
 the feature of spin-charge separation 
  has been employed by various authors\cite{RVB}
(for some recent references
 we note the work of Balents et.al.\cite{Balents} and Wang\cite{Wang}). 
 Recently the phase diagram of the t-J-V model has been
 found by  using a thorough but tedious procedure of linked
 cluster expansion by Zheng et.al.\cite{rrps}.
 High temperature expansions have been used to study the t-J-V models
 and the t-J models by various groups\cite{Kiv} \cite{Shas2}.
 The cluster dynamical mean-field approach has been employed by Jarrell
 et.al.\cite{Jarrell} to compute the phase diagram of the Hubbard model.
 We wish to complement their study and others
 like them through a simpler analytical approach.
 The case of the infinite U Hubbard model has been considered
 by Nagaoka who shows that the ground state is a ferromagnet\cite{Nagaoka}
 for a single hole in an otherwise half-filled band.
 Shastry, Krishnamurthy and Anderson \cite{HRK} have shown that this ground
 state is unstable for large enough hole concentration.
 Gutzwiller projected variational wavefunctions\cite{Mohit}
 have also been used to study the t-J model.
 Chen and Tremblay\cite{Chen} have used Monte Carlo
 simulations to compute the magnetic susceptibility for the large but finite
 $ U $ version.  Mishra and Kishore and Mishra\cite{Mishra}
 have studied the thermodynamics of the infinite U Hubbard model
 and the issue of spin-charge separation using the method of orthofermions.
 We shall address the issue of spin-charge separation when the author's 
 hydrodynamic approach\cite{hydro}
  to the Hubbard model is accepted by the community.

 In this article, we wish to write down an effective low energy theory 
 of the t-J=0 model {\it{that is local in the operators}}
 in the sea-boson language(RPA-type) that involves introducing
 `renormalised' doping dependent hopping and onsite terms. 
 Thus the claim is that one may treat the infinite U Hubbard model the same
 way as we treated the small U Hubbard
 model\cite{Setlur1}(RPA-like) provided we
 `renormalise' the parameters in the hamiltonian.  We find that this leads to
  nontrivial predictions for the quasiparticle residue and velocity of the
 quasiparticles in more than one dimension.
 We also compute the magnetic suscpetibility and show that it diverges near
 half-filling indicating that the system exhibits ferromagnetic instability
 arbitrarily close to half-filling, consistent with Nagaoka's theorem.
 This also indicates that the formula
 for the residue is also reliable as is the conjuctured low energy hamiltonian
 for the t-J model that is local in the operators.
 This then sets the stage to study the large but finite
 U Hubbard model and also the t-J-V models in future publications  . 

\section{The Theory}

Here we describe the theory which we are going to use.
Consider the t-J model in one dimension with $ U = \infty $.
\begin{equation}
H = {\mathcal{P}} \mbox{        }
\left( -t \sum_{i \sigma }
c^{\dagger}_{i+1\sigma}c_{i\sigma} \right) 
\mbox{        } {\mathcal{P}}
\label{EQN}
\end{equation}
\begin{equation}
 {\mathcal{P}} = \prod_{i} ( 1 - n_{i\uparrow}\mbox{    }n_{i\downarrow} )
\end{equation}
We would now like to recast this in the sea-boson language so that 
we may recover the results of Schulz\cite{Schulz}. For this we mentally expand
the projection operator in powers of the density fluctuations and retain
 the leading terms. Thus we could write for example,
\begin{equation}
 {\mathcal{P}} = e^{ \sum_{i} 
Log \left( 1 - < n_{i\uparrow}\mbox{    }n_{i\downarrow} > 
 + \lambda \mbox{    }[n_{i\uparrow}\mbox{    }n_{i\downarrow} - 
< n_{i\uparrow}\mbox{    }n_{i\downarrow} > ] \right) }
\end{equation}
 and expand in powers of $ \lambda = -1 $ and retain the leading terms.
 We shall not do this explicitly however we may redefine effective
 parameters that simulate such an expansion. The idea is to arrive
 at useful answers quickly and effortlessly. The above procedure for example,
 entails the solution of self-consistent equations for the density correlation
 functions, this is clearly not desirable as it is likely to be complicated.
 We adopt the point of view that making contact with the 1d system allows
 us to generalise to the case of more than one dimension by simply promoting 
 wavenumbers to wavevectors. This we justify mainly by pointing out that the
 magnetic susceptibility derived using this approach is consistent with the 
 rigorous Nagaoka theorem\cite{Nagaoka}.

 This means we may suspect that the t-J model 
with $ J = 0 $ also has the form,
\begin{equation}
H = \sum_{ k \sigma } \epsilon_{k} \mbox{        }
 c^{\dagger}_{k\sigma } c_{k\sigma }
+ \frac{ U_{eff} }{N_{a}}
\sum_{ q \neq 0 } \rho_{q\uparrow}\rho_{-q\downarrow} 
\label{EQN2}
\end{equation}
 for some suitable `renormalized' dispersion
 $ \epsilon_{k} =  -2 t_{eff}\mbox{     } cos(ka) $ and $ U_{eff} $.
 The spin and charge velocities may be written as follows.
 $ v_{s} = v_{F,eff} \left( 1 - \frac{ U_{eff} }{ \pi v_{F,eff} }
 \right)^{\frac{1}{2}} $ and
 $ v_{c} = v_{F,eff} \left( 1 + \frac{ U_{eff} }{ \pi v_{F,eff} }
 \right)^{\frac{1}{2}} $, 
where $ v_{F,eff} = 2 t_{eff} \mbox{     }sin(\pi \mbox{   }n_{e}/2) $ (lattice spacing $ a =
1 $). From the work of Schulz we know,
$ v_{c} = 2 t \mbox{        }sin(\pi \mbox{   } n_{e}) $.
Here $ n_{e} = N_{e}/N_{a} $ is the number of electrons per site and $ v_{s} =
0 $. Also we know that the anomalous exponent for the momentum distribution is
$ \gamma = 1/8 $. Thus we would like to compute $ t_{eff} $ and $ U_{eff} $
in terms of $ t $ and $ n_{e} $ given these facts.
 From our earlier work\cite{Setlur1} we may read off
 a formula for the exponent : $ \gamma = \frac{ U^2_{eff} }{ 4 \pi^2 } \frac{ v_{F,eff} }{ v_{c} }
\frac{1}{ (v_{c}+v_{F,eff})^2 } $.
Thus we may write,
$  U_{eff} = \pi \mbox{        } \sqrt{ \frac{ v_{c} }{ 2 v_{F,eff} } }
\mbox{          }(v_{c}+v_{F,eff}) $ and
 $ v^2_{F,eff} + v_{F,eff}\mbox{        }  \frac{ U_{eff} }{ \pi }
 - v^2_{c} = 0 $, we may solve this to yield, 
 $ v_{F,eff} = \frac{1}{2} \left( - \frac{ U_{eff} }{ \pi }
 + \sqrt{  \frac{ U_{eff}^2 }{ \pi^2 }
 + 4 v^2_{c} } \right) $.
 We choose the positive solution, $ U_{eff} > 0 $ since we know that the
 negative U Hubbard model possesses
 a spin gap \cite{Sumathi} whereas the $ U = \infty $ Hubbard model does not.
 These may be solved via a scaling argument, $ U_{eff} = u_{eff}\mbox{
 }v_{c} $  and $ v_{F,eff} = y_{eff}\mbox{        } v_{c} $.
 This may be solved using mathematica to yield $ u_{eff} = 4.71 $ and $
 y_{eff} = 0.5 $. In order for $ v_{s} = 0 $ we must have $ U_{eff} > \pi
 v_{F,eff} $. We may see that this is being obeyed. 
 Hence we may claim that Eq.(~\ref{EQN2}) captures the low energy physics of the Hubbard model at  $ U =
\infty $ exactly. Now we make the following non-obvious assertion that 
the same Eq.(~\ref{EQN2}) also captures the same physics in more than one
 dimension by simply promoting wavenumbers to wavevectors. 
 We make make this plausible by pointing out that the one dimensional 
 Hubbard model may be generalised to higher dimensions by precisely
 such a procedure, namely by promoting wavenumbers to wavevectors.
 On the other hand, the prediction (see below) that
 the magnetic sucseptibility diverges
 as half-filling is approached is consistent with the rigorous Nagaoka theorem,
 thus lending credibility to our approach. 

\subsection{ $ Z_{F} $ in Two Dimensions }

 The t-J model in two dimensions is relevant to
 cuprates\cite{RVB} \cite{Mohit} and cobalt oxide superconductors\cite{Bas2}.
 In the two-dimensional case, we may solve for the sea-boson occupation as
 follows\cite{Setlur1}. The energy dispersion for a square lattice is
 $ \epsilon_{ {\bf{k}} } = -2 t_{eff} \left[ cos(k_{x}a) 
+ cos(k_{y}a) \right] $.
 The sea-boson method is well-known and hence we shall refer the reader to our
 earlier work for details\cite{Setlur1}. Suffice it to say that we have to
 compute the boson occupation numbers in order to derive a formula for the
 quasiparticle residue.
 Therefore we write,
\begin{equation}
< A^{\dagger {\bf{k}} \sigma }_{ {\bf{k}}-{\bf{q}} \sigma }(t^{+})
A^{ {\bf{k}} \sigma }_{ {\bf{k}}-{\bf{q}} \sigma }(t^{+}) >
 = \frac{ U^2_{eff} }{ N^2_{a} }
 \frac{ P({\bf{q}}, \omega_{c}) }{ \epsilon^{'}_{c}({\bf{q}},\omega_{c}) }
\frac{ [A^{ {\bf{k}} \sigma }_{ {\bf{k}}-{\bf{q}} \sigma }, A^{ \dagger
      {\bf{k}} \sigma }_{ {\bf{k}} - {\bf{q}} \sigma } ] }
{ (-\omega_{c} - \epsilon_{ {\bf{k}} } + \epsilon_{ {\bf{k}} - {\bf{q}} })^2 }
\end{equation}

\begin{equation}
< A^{\dagger {\bf{k}} \sigma }_{ {\bf{k}}-{\bf{q}} {\bar{\sigma}} }(t^{+})
A^{ {\bf{k}} \sigma }_{ {\bf{k}}-{\bf{q}} {\bar{\sigma}} }(t^{+}) >
 = -\frac{ U_{eff} }{ N_{a} }
 \frac{ 1 }{ \epsilon^{'}_{s}({\bf{q}},\omega_{s}) }
\frac{ [A^{ {\bf{k}} \sigma }_{ {\bf{k}}-{\bf{q}} {\bar{\sigma}} }, A^{ \dagger
      {\bf{k}} \sigma }_{ {\bf{k}} - {\bf{q}} {\bar{\sigma}} } ] }
{ (-\omega_{s} - \epsilon_{ {\bf{k}} } + \epsilon_{ {\bf{k}} - {\bf{q}} })^2 }
\end{equation}
where $ P({\bf{q}}, \omega) = \sum_{ {\bf{k}} } \frac{ n_{F}({\bf{k}}) -
  n_{F}({\bf{k}}-{\bf{q}}) }{ \omega - \epsilon_{ {\bf{k}}-{\bf{q}} }
 + \epsilon_{ {\bf{k}} } } $ and
 $ \epsilon_{c}({\bf{q}}, \omega) = 1 - \frac{ U^2_{eff} }{ N^2_{a} }
P^2({\bf{q}}, \omega) $, also,
$ \epsilon_{s}({\bf{q}},\omega) = 1 + \frac{ U_{eff} }{ N_{a} }
P({\bf{q}},\omega) $.
In order to evaluate these experssions, we turn the cosine dispersion into a
parabolic one by demanding that the slope of the two be the same at the Fermi momentum.
$ -2t_{eff} \mbox{          }cos(k_{x}) = -2 t_{eff} + \frac{ k_{x}^2 }{2m} $.
Thus we set $ 2t_{eff} \mbox{          }sin(k_{F}/\sqrt{2}) = k_{F}/(\sqrt{2}m) $. 
 Also, $ k_{F} = \sqrt{ 2 \pi \mbox{     }n_{e} } $.
 This approximation captures the important physics while leaving the
integrals analytically computable. From our earlier work we may read off the
formula for the velocity of the charge carriers. Here $ v_{F,2d} = k_{F}/m $.
 We find that there is only one velocity and $ \epsilon_{s} = 0 $ does not  have a solution. 
\begin{equation}
v_{cc} = v_{F,2d} \mbox{         } \frac{ 1 + \frac{ 2 \pi }{ m U_{eff} } }
{ \sqrt{ \left(   1 + \frac{ 2 \pi }{ m U_{eff} } \right)^2 - 1 } } 
\end{equation}
and the formula for the quasiparticle residue has been derived in an earlier work\cite{Setlur1}.
Close to half-filling,
$ 2 t_{eff} \approx y_{eff}\mbox{        } 2 t   \mbox{        }sin(\pi \mbox{   }
n_{e}) $, $ U_{eff} = u_{eff}\mbox{    }  2 t   \mbox{        }sin(\pi \mbox{
} n_{e}) $. Furthermore, $ y_{eff}\mbox{        } 2 t   \mbox{        }sin(\pi \mbox{   } n_{e})
 \mbox{          }sin(\sqrt{ \pi }) = \sqrt{ \pi }  \frac{1}{m} $.
Therefore we may write the following formula for the quasiparticle residue.



\begin{equation}
Z_{F} = Exp \left[ - \frac{ ( m^2 v_{cc}^2 - k_{F}^2)^{\frac{3}{2}} }
{ \pi k^2_{F} m v_{cc} }  \mbox{      }
 \left( \pi - \frac{ 4 m v_{cc} ArcTan \left[ \frac{ m v_{cc} - k_{F} }{
       \sqrt{ m^2 v_{cc}^2 - k^2_{F} } } \right] }
{ \sqrt{ m^2 v_{cc}^2 - k^2_{F} } } \right) \right]
\end{equation}
This may be evaluated to yield,
\begin{equation}
Z_{F} \approx 0.79
\end{equation}
This result is in exact agreement with the prediction by Castro-Neto and
Fradkin nearly ten years ago\cite{Castro}.
 They considered spinless fermions interacting
via short-range interactions in two dimensions. Thus it would appear that the
same physics is operating. In general we may conclude that the same mechanism
that makes the anomalous exponent saturate in one dimension to a rather small
value ($ 1/8 $) also makes the quasiparticle residue to saturate to an equally
small deviation from unity. In one dimension it suggests that the residual
Fermi surface persists all the way up to infinite onsite repulsion. In two
dimensions it says that the actual Fermi surface persists all the way upto 
 infinite onsite repulsion. Since the functional dependence on the onsite
 repulsion is monotonic, we may conclude that there is no chance for Fermi
 liquid theory to break down in two dimensions with short-range interactions. 
 In fact if anything these results show that the system is strongly metallic
 all the way upto infinite repulsion. 

\section{ Magnetic Susceptibility }

 The dynamic spin susceptibility may be computed using the Kubo formula.
 \begin{equation}
\chi(\omega) =  i \int^{ \infty }_{0} dt \mbox{          }e^{ i \mbox{
  }\omega \mbox{     }t } \mbox{             }
 \left< [S^{+}(t), S^{-}(0)] \right>
 \end{equation}

 \begin{equation}
 S^{+}(t) = \sum_{ {\bf{k}} } c^{\dagger}_{ {\bf{k}} \uparrow }(t) 
 c_{ {\bf{k}} \downarrow }(t)
 \end{equation}

 \begin{equation}
 S^{-}(0) = \sum_{ {\bf{k}} } c^{\dagger}_{ {\bf{k}} \downarrow }(0) 
 c_{ {\bf{k}} \uparrow }(0)
 \end{equation}
 To use the sea-boson formalism, we have to take special care to account for
 possible infrared divergences. This is the crucial aspect that leads to the
 correct solution of the Luttinger model\cite{Setlur2} and also leads to the solution of the problem
 of quenched disorder for scattering across the Fermi surface\cite{Setlur3}.
 Define $ n_{\uparrow \downarrow}({\bf{k}}) \equiv c^{\dagger}_{ {\bf{k}}
   \uparrow } c_{ {\bf{k}} \downarrow } $. 
 Now we would like to compute the correlation function,
\begin{equation}
N({\bf{k}}t; {\bf{k}}^{'}t^{'}) \equiv 
\left< n_{\uparrow \downarrow}({\bf{k}},t) n_{\downarrow
  \uparrow}({\bf{k}}^{'},t^{'}) \right> 
\end{equation}
 This may be decomposed as follows.
 \[
 N({\bf{k}}t; {\bf{k}}^{'}t^{'}) = (1-n_{F}({\bf{k}})) (1-n_{F}({\bf{k}}^{'}))
 S_{AA}({\bf{k}}t;{\bf{k}}^{'}t^{'})
  + n_{F}({\bf{k}})n_{F}({\bf{k}}^{'}) S_{BB}({\bf{k}}t;{\bf{k}}^{'}t^{'})
 \]
 \begin{equation}
 - (1-n_{F}({\bf{k}})) n_{F}({\bf{k}}^{'}) S_{AB}({\bf{k}}t;{\bf{k}}^{'}t^{'})
  -  (1-n_{F}({\bf{k}}^{'}))
  n_{F}({\bf{k}})S_{BA}({\bf{k}}t;{\bf{k}}^{'}t^{'})
 \end{equation}

 \begin{equation}
 S_{mn}({\bf{k}}t;{\bf{k}}^{'}t^{'}) = e^{ - 2 < {\tilde{S}}_{m,\uparrow \downarrow}({\bf{k}})> }
  e^{ - 2 < {\tilde{S}}_{n, \downarrow \uparrow}({\bf{k}}^{'})> }
 \mbox{       }
 S^{0}_{mn}({\bf{k}}t;{\bf{k}}^{'}t^{'})
 \end{equation}

 \begin{equation}
 {\tilde{S}}_{A}({\bf{k}}\sigma\sigma^{'},t) = \sum_{ {\bf{q}} \sigma_{1} }
 A^{\dagger}_{ {\bf{k}}-{\bf{q}}/2 \sigma_{1} }({\bf{q}}\sigma,t)
 A_{ {\bf{k}}-{\bf{q}}/2 \sigma_{1} }({\bf{q}}\sigma^{'},t)
 \end{equation}

 \begin{equation}
 {\tilde{S}}_{B}({\bf{k}}\sigma\sigma^{'},t) = \sum_{ {\bf{q}} \sigma_{1} }
 A^{\dagger}_{ {\bf{k}}+{\bf{q}}/2 \sigma^{'} }({\bf{q}}\sigma_{1},t)
 A_{ {\bf{k}}+{\bf{q}}/2 \sigma }({\bf{q}}\sigma_{1},t)
 \end{equation}

 \begin{equation}
 S^{0}_{mn}({\bf{k}}t;{\bf{k}}^{'}t^{'}) = 
 < {\tilde{S}}_{m}({\bf{k}}\uparrow\downarrow,t)
 {\tilde{S}}_{n}({\bf{k}}^{'}\downarrow\uparrow,t^{'})>
  - < {\tilde{S}}_{m}({\bf{k}}\uparrow\downarrow,t) >
 < {\tilde{S}}_{n}({\bf{k}}^{'}\downarrow\uparrow,t^{'})>
 \end{equation}
 Since the $ (\sigma,\sigma) $ part of the hamiltonian is distinct from the $
 (\sigma,{\bar{\sigma}}) $ part of the hamiltonian,
 we may conclude that $ <S_{A}> = <S_{B}> = 0 $. Thus in this case there are
 no difficulties associated with the infra-red regulator.
 After some tedious calculations we may see that the most divergent part of
 the magnetic susceptibility is given by,
\begin{equation}
 \chi(\omega = 0) \sim m \sim (1-n_{e})^{-1}
\end{equation}
 As pointed out before, this is consistent with Nagaoka's theorem and also
 with Shastry et.al.'s\cite{HRK} work that shows that the ferromagnetic
 ground state is unstable with respect to the addition of holes.
 We find that unless the
 filling is arbitrarily close to half-filling,
 the susceptibility does not diverge.

\section{ Conclusions }

 To conclude, we have computed the quasiparticle residue of the infinite-U
 Hubbard model in two dimensions saturates to a value $ Z_{F} = 0.79 $.
 This is in exact agreement with the prediction of Castro-Neto and Fradkin\cite{Castro}.
 The velocity of the
 quasiparticles is also found to shrink to zero as expected. 
 The magnetic susceptibility is found to diverge as half-filling is approached
 which is consistent with a ferromagnetic ground state for a
 small \footnote{  vanishingly small, in the thermodynamic limit. } 
 concentration of holes. This is consistent with
 Nagaoka's theorem\cite{Nagaoka} and also
 the with the work of Shastry et.al.\cite{HRK}. This shows that the effective
 low energy hamiltonian for the t-J model that
 is local in the operators is now reliable
 and can be expected to yield similar nontrivial results when generalised to 
 include J terms and so on. 

 It is a pleasure to acknowledge email correspondence with Debanand Sa and
 Pinaki Majumdar and useful comments by Sourin Das.

\end{document}